\documentclass[twocolumn,showpacs,preprintnumbers,amsmath,amssymb]{revtex4}

\usepackage{graphicx}
\usepackage{dcolumn}
\usepackage{bm}

\begin{document}
\title{Nernst effect in semi-metals: the meritorious heaviness of electrons}
\author{Kamran Behnia$^{1}$, Marie-Aude M\'easson$^{2}$ and Yakov Kopelevich$^{3}$}
\affiliation{(1)Laboratoire de Physique Quantique(CNRS), ESPCI, 10
Rue de Vauquelin,
75231 Paris, France \\
(2)Graduate School of Science, Osaka University, Toyonaka, Osaka,
560-0043 Japan \\
(3)Instituto de Fisica ``Gleb Wataghin'', UNICAMP, 13083-970
Campinas, S\~{a}o Paulo, Brazil}

\date {December 14, 2006}

\begin{abstract}
We present a study of electric, thermal and thermoelectric transport
in elemental Bismuth, which presents a Nernst coefficient much
larger than what was found in correlated metals. We argue that this
is due to the combination of an exceptionally low carrier density
with a very long electronic mean-free-path. The low thermomagnetic
figure of merit is traced to the lightness of electrons.
Heavy-electron semi-metals, which keep a metallic behavior in
presence of a magnetic field, emerge as promising candidates for
thermomagnetic cooling at low temperatures.
\end{abstract}


\maketitle

In presence of a  magnetic field , the application of a thermal
gradient to a solid may generate an electric field orthogonal to
both of them. This is the Nernst effect that was discovered by
Ettingshausen and Nernst 120 years ago in a study of elemental
Bismuth\cite{ettingshausen}. During the last few years, following
the observation of a finite Nernst signal in the normal state of the
underdoped cuprates\cite{xu}, this effect was studied for the first
time in metals host to correlated electrons. In several cases, a
large Nernst signal was unexpectedly
resolved\cite{wu,bel1,bel2,choi,sheikin,pourret,nam}. Theoretically,
the Nernst response of a simple metal is expected to vanish in the
absence of electron-hole asymmetry\cite{sondheimer,wang}. An
``ambipolar'' Nernst effect is present when the metal is
compensated\cite{bel3}, but the magnitude of the ``giant'' Nernst
effect observed in correlated metals remained puzzling. In most
cases, this signal emerged when the system was apparently out of the
realm of the Fermi liquid picture. In CeCoIn$_5$, the large Nernst
signal was concomitant with anomalous behavior in various electronic
properties of the system\cite{bel1}. In the Bechgaard salts, it
occurred when the field was oriented close to the magic Lebed
angles\cite{wu,choi}. In URu$_{2}$Si$_{2}$\cite{bel2} and
PrFe$_{4}$P$_{12}$\cite{pourret}, the giant signal emerged with the
establishment of exotic electronic orders. These observations
suggested a possible link between a large Nernst signal and
non-Fermi liquid physics and raised a fundamental question : what
sets the magnitude of the Nernst response of a Fermi liquid ?

\begin{figure}
\resizebox{!}{0.6\textwidth} {\includegraphics{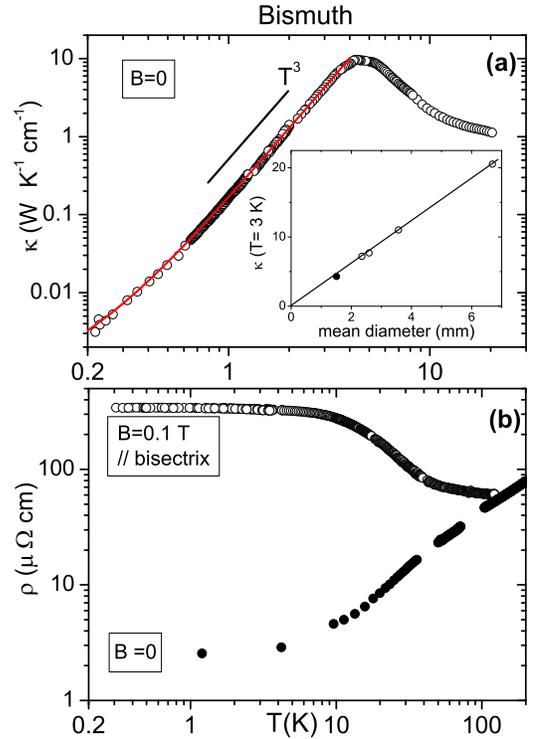}}x
\caption{\label{fig1} (a) Thermal conductivity, $\kappa$  of the Bi
single crystal. Solid line represents a $aT+bT^{3}$ fit (see text).
Inset compares the magnitude of $\kappa(3K)$ of the sample of this
study(solid circle) with those reported in ref\cite{boxus} (open
circles) as a function of mean diameter. (b) Resistivity of the same
sample at zero field and in presence of a field of 0.1 T. }
\end{figure}
In this paper, we present a study of thermal and thermoelectric
transport in elemental Bismuth down to 0.2K. Our study confirms that
in this semi-metal the Nernst coefficient exceeds by an order of
magnitude the largest signal observed in correlated metals. We argue
that the Nernst coefficient of a Fermi liquid roughly tracks
$\omega_{c}\tau/\epsilon_{F}$; with $\omega_{c}$, the cyclotron
frequency, $\tau$, the scattering time and $\epsilon_{F}$, the Fermi
energy. The exceptionally low value of carrier density in Bismuth
combined to a very long electronic mean-free-path in clean single
crystals is a source of giant Nernst signal. However, because of the
low effective mass of the quasi-particles, the electric conductivity
is easily degraded by the application of a magnetic field. A
remarkable difference between light- and heavy-electron semi-metals
is that only the latter continue to behave like a metal in presence
of a moderate magnetic field and could be used for constructing a
cryogenic Ettingshausen refrigerator.

Figure 1 presents the thermal conductivity, $\kappa$ and electric
resistivity, $\rho$, of the Bi single crystal (dimensions: 2.2
$\times$ 1.1 $\times$ 0.8 mm$^{3}$) used in this study. In all
measurements, heat or charge was injected along the binary axis and
the magnetic field was oriented either along the trigonal or the
bisectrix axes. As seen in the lower panel, the residual resistivity
was $\rho_{0} \sim 2.5 \mu \Omega$ cm (i.e. RRR = 47). The thermal
conductivity, $\kappa$, displayed in the upper panel, presents a
maximum at 4.1 K. Below this temperature, $\kappa$ follows a $T^3$
behavior characteristic of $ballistic$ lattice thermal conductivity.
The electronic contribution, which generates a small $T$-linear term
becomes only visible below 0.5 K. The expression $\kappa=aT+bT^{3}$
with $a = 1 WK^{-2}m^{-1}$ and $b= 158 WK^{-2}m^{-1}$ fits the data
up to 3.6 K. The first term represents the electronic contribution
and, as expected by the Wiedemann-Franz law: $a\simeq
L_{0}/\rho_{0}$ (L$_{0}$ is the Lorenz number). The second term
represents the lattice thermal conductivity. Since,
$\kappa_{ph}=\frac{1}{3}C_{ph}v_{s}\ell_{ph}$, by taking the
reported values for the sound velocity, $v_{s} = 1100 m/s
$\cite{kopylov}, the lattice specific heat ($C_{ph}=35 T^{3} J
m^{-3}K^{-1}$\cite{collan}) and the measured value of $b$, one can
estimate $\ell_{ph}$ = 1.1mm. The closeness of this length to the
sample's mean diameter (1.5mm) validates the hypothesis of ballistic
phonon transport. Our data should also be compared with a previous
study of thermal conductivity (restricted to T$>2$K) on crystals of
various dimensions. \cite{boxus}. Above 4 K, the phonon
mean-free-path drastically decreases as a function of temperature,
$\kappa$ is not set by the sample size and our data can be
superposed on the results reported by Boxus \emph{et
al.}\cite{boxus}. Below 4 K, and as seen in the inset of the figure
which compares our data with the samples used in that study,
$\kappa$ at a given temperature (say 3 K) is simply proportional to
the sample's mean diameter. This provides further evidence that the
phonon mean-free-path is set by the sample size.

The lower panel of Fig. 1 recalls the remarkably large
magneto-resistance of Bismuth. The application of a modest magnetic
field of 0.1 T enhances the magnitude of resistivity by more than
two orders of magnitude. In Bismuth as well as in
graphite\cite{kopelevich,du}, the magnetic field induces an
insulating-like behavior. The ultimate criterion to qualify as a
metal, however, is to have a Fermi surface and this is the case of
the system under study in the zero-temperature limit. The giant
magnetoresistance can be traced to the large value of
$\omega_c\tau$. In Bismuth, there are  $3\times 10^{17}$ holes per
cm$^3$ and a same density of electrons. Therefore, the magnitude of
$\rho_{0}$ implies $\omega_c\tau=\frac{eB \tau}{m^{*}}\simeq 42$ at
0.1 T and a 300-fold increase in resistivity is
unsurprising\cite{du}. The opening of an excitonic insulating gap at
low applied magnetic fields has also been suggested
\cite{kopelevich2}.

\begin{figure}
\resizebox{!}{0.35\textwidth} {\includegraphics{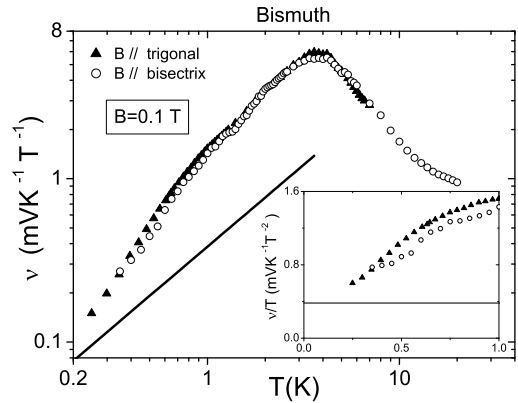}}
\caption{\label{fig2} The temperature dependence of the absolute
value of the Nernst coefficient of the Bismuth single crystal for
two different orientations of the magnetic field.  The solid line
represents a linear function $\alpha$T with
$\alpha=283\frac{\omega_{c}\tau}{\epsilon_{F}B}=0.38mVK^{-2}T^{-1}$
(see text and table 1). Both this function and the low-temperature
data are displayed in the inset as a $\nu/T$ vs. T plot.}
\end{figure}

We now turn to the Nernst coefficient. Fig. 2 presents the
temperature dependence of $|\nu|=N/B=E_{y}/(B \nabla_{x}T)$. As seen
in the figure, we found that for two orientations of the magnetic
field $\nu$ peaks at 3.8 K to a value of 7 $mV/KT$. This large value
falls in the range of magnitudes reported in studies published
decades ago\cite{sugihara,korenblit,mangez}.

Since the contribution of electrons to heat transport is negligible
and since bismuth is a compensated metal ($n_{e} \simeq n_{h}$) with
a Hall angle much smaller than $\omega_{c}\tau$, there is no
surprise that we did not detect any measurable thermal Hall effect.
As  heat current and temperature gradient vectors remain parallel in
presence of magnetic field, the adiabatic and the isothermal Nernst
coefficients are virtually identical in bismuth\cite{sugihara}.

As seen in Figure 3, the magnitude of the Nernst coefficient in
Bismuth is such that it dwarfs what is reported for other metals,
even  those subject to a generous attribution of the adjective
``giant''. It is generally accepted that Bismuth is a Fermi liquid.
Why then the magnitude of its Nernst coefficient is so large? We
will argue below that this is because of its unique electronic
properties\cite{edelman}, namely the combination of an exceptionally
low carrier density ($10^{-5}$ carriers per atom) and a very long
electronic mean-free-path (40 $\mu$m in our sample).

Before this, let us briefly consider the role played by phonons. In
Bismuth, around  3 K, the typical phonon wave-vector becomes
comparable to 2$k_{F}$\cite{uher}. Therefore, phonon drag should be
the most important source for the Nernst signal at its peak
temperature\cite{sugihara,korenblit}. Conceptually, it is easier to
picture this phenomenon\cite{nolas} in an Ettingshausen geometry:
when electrons loose their impulsion in a collision with phonons,
the electric current gives rise to  an  entropy current of phononic
origin, hence a finite Ettingshausen effect. Since the Bridgemen
relation ties the amplitudes of the Ettingshausen and Nernst
effects\cite{nolas}, this implies that the Nernst effect should also
be enhanced. In our case, below 4K, electron-phonon scattering does
not manifest itself either in charge or heat transport. Ballistic
phonon conductivity means that the electrons do not scatter phonons
in a visible way. As for charge conductivity, it changes by less
than 10 percent below 4K, which means that the electrons are also
mostly scattered by defects. All this indicates that electron-phonon
scattering events in this regime do not occur frequently. Even
though, since the phonon lifetime is orders of magnitude longer than
the inelastic lifetime of electrons, such rare events can
drastically amplify the Nernst effect. The presence of this
phonon-drag Nernst effect complicates the quantitative analysis of
the purely electronic (often called diffusive) component of the
Nernst effect. However, a finite $\nu/T$ persists down to the lowest
temperatures of this study (see the inset of the figure). Let us
argue that the magnitude of this T-linear $\nu$ is comparable to
what is expected from the electronic properties of Bismuth.

Within the Boltzmann picture, the Nernst coefficient of a metal is
given by the following expression\cite{clayhold,oganesyan} :

\begin{equation}\label{2}
\nu= \frac{\pi^{2}}{3}\frac{k_{B}^{2}T}{B e}\frac{\partial
\tan\theta_{H}}{\partial \epsilon}|_{ \epsilon_{F}}
=\frac{\pi^{2}}{3}\frac{k_{B}^{2}T}{m^{*}}\frac{\partial
\tau(\epsilon)}{\partial \epsilon}|_{ \epsilon_{F}}
\end{equation}

which was first derived by Sondheimer\cite{sondheimer}. The first
expression  links $\nu$ to the Hall angle, $\tan\theta_{H}$  and was
discussed in detail in ref. \cite{oganesyan}. In the case of
Bismuth, because of the compensation between electrons and holes,
$\tan\theta_{H}$ is much smaller than $\omega_{c}\tau$ of each type
of carriers. In order to estimate the order of magnitude of the
Nernst coefficient in a metal, let us replace $ \frac{\partial
\tau}{\partial \epsilon}|_{ \epsilon_{F}}$ by
$\frac{\tau}{\epsilon_{F}}$\cite{bel2}. This leads to the following
gross scale for the Nernst signal:
\begin{equation}\label{3}
\nu \approx 283 \frac{\mu V }{K} \times \frac{\omega_{c}\tau}{B}
\times \frac{k_{B}T}{\epsilon_F}
\end{equation}

Thus, there are three distinct ways of enlarging a $T$-linear $\nu$:
increasing the scattering time, increasing the cyclotron Frequency
and reducing the Fermi energy. All these three routes are taken in
Bismuth. The samples are clean (enhancing $\tau$), the effective
mass is small (leading to a large $\omega_{c}$) and finally (and
most importantly) the carrier density is low (pushing down the Fermi
energy in spite of the reduced effective mass). The solid line in
Fig. 2 represents what is expected according to this expression
taking $\omega_{c}\tau/B =420T^{-1}$ and $\epsilon_{F}/k_{B}$ =310
K\cite{liu}. It appears to give a satisfactory account of the purely
diffusive part of the Nernst signal.

\begin{table}
\begin{tabular}{|c|c|c|c|c|}
  \hline
    & Bi  &  PrFe$_{4}$P$_{12}$ & URu$_{2}$Si$_{2}$ & Ref.  \\
  \hline
  $k_{F}$(nm$^{-1}$) & 0.14  & 0.7 & 1.1 & \cite{edelman,sugawara,ohkuni} \\
  \hline
 $m^{*} (m_{e})$ & 0.06 & 10&  25& \cite{liu,sugawara,ohkuni} \\
    \hline
$\gamma (mJK^{-1}mol^{-1})$ &0.048  & 100 & 65 &\cite{collan,aoki,maple} \\
    \hline
$n$ (per f.u.) & $10^{-5}$  & 5-18$\times 10^{-3}$ &3-5$\times 10^{-2}$ &\cite{edelman,sugawara,pourret,bel2}\\
    \hline
$\ell_{e}$ ($\mu$m) & 40 & 0.4 & 0.1 & \cite{pourret,bel2}  \\
    \hline
$\omega_c \tau$(B=1T) &420 & 0.85 &  0.08&  \\
    \hline
$\epsilon_{F}$(K)&310 & 9 &22&  \cite{liu,aoki} \\
  \hline
283 $\frac{\omega_{c}\tau}{\epsilon_{F}}$($\mu$V/K$^{2}$T)&383 & 27 &1.1&   \\
  \hline
$\nu /T$($\mu$V/K$^{2}$T)(0.3 K) & 750   & 57  & 2.4 &\cite{pourret,bel2}   \\
    \hline
$ZT_{\epsilon}$(1K)& $< 0.001$ & 0.19 & $\sim0.01$&\cite{pourret,bel2,behnia}   \\
   \hline
\end{tabular}

\caption{A comparison of three semi-metals. $k_{F}$ and $m^{*}$ are
the radius and the effective mass of a Fermi surface (the hole
ellipsoid s in Bismuth, the $\tau$ and the $\beta$ bands in the
ordered states of PrFe$_{4}$P$_{12}$ and URu$_{2}$Si$_{2}$).
$\gamma$ and $n$ are electronic specific heat and carrier density.
The electronic mean-free-path, $\ell_{e}$, is estimated for samples
used in the Nernst studies. The value for $ZT_{\epsilon}$ is the
highest obtained for $B <12 T$. } \label{T1}
\end{table}

Let us now compare Bismuth with URu$_2$Si$_2$ and
PrFe$_{4}$P$_{12}$. In the two heavy-fermion metals, a large Nernst
signal emerges below a temperature associated with an exotic phase
transition. The case of Bismuth suggests that the most likely source
of the large Nernst signal in these two compounds is the
semi-metallic nature of the ordered system after the opening of a
gap which destroys most of the Fermi surface. Table 1 gives a list
of electronic properties of these three metals. The low level of the
carrier density is one feature that they share. As seen in the
table, the magnitude of the Nernst effect scales with
$\omega_{c}\tau/\epsilon_{F}$. This conclusion implies that future
experiments should resolve a large Nernst signal in heavy-fermion
semi-metals \emph{without} any exotic order. While the case of
Bismuth shows that a giant Nernst effect cannot be considered as a
solid signature of non-Fermi liquid physics, it also demonstrates
that we still lack a precise quantitative understanding of the
Nernst coefficient even in the zero-temperature limit.

\begin{figure}
\resizebox{!}{0.35\textwidth} {\includegraphics{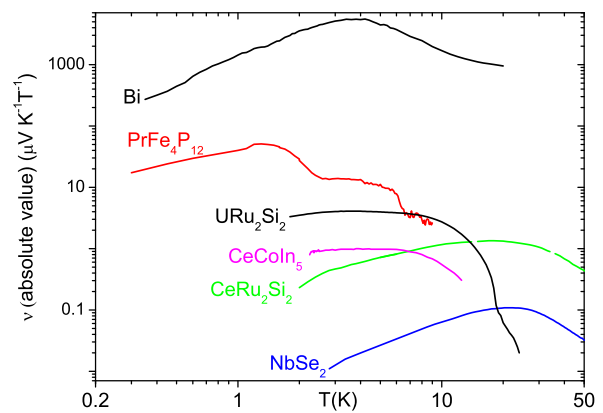}}
\caption{\label{fig3} The magnitude of the Nernst coefficient in
Bismuth compared to what is found in some other
metals\cite{bel1,bel2,sheikin,pourret,bel3}.}
\end{figure}

The result has also an implication for applied research. In the
quest for useful thermoelectricity\cite{nolas,mahan}, the figure of
merit, ZT= $\frac{S^{2}\sigma T}{\kappa}$, quantifies the adequacy
of a given material for thermoelectric refrigeration. In metals, the
Wiedemann-Franz law sets the magnitude of $\frac{\kappa}{\sigma T}=
L_{0}=2.44 \times10^{-8} V^{2}/K^{2}$. Therefore, a thermopower of
$S\approx \sqrt{L_{0}}=155 \mu V/K$ would imply  $ZT \approx 1 $.
Recently, Harutyunyan \emph{et al.}\cite{harutyunyan} used CeB$_{6}$
(with $S\simeq120 \mu V/K$ around 6 K) to construct a Peltier cooler
at cryogenic temperatures\cite{harutyunyan}. In the case of an
Ettingshausen refrigerator, the relevant parameter is the
thermomagnetic figure of merit $ZT_{\epsilon}= \frac{N^{2}\sigma
T}{\kappa}$\cite{nolas} and $\sqrt{L_{0}}$ sets a similar threshold
for $N$. Therefore, its magnitude in PrFe$_{4}$P$_{12}$ ($\sim100
\mu V/K$ at T=1K and B=4T) opens a possible route for thermomagnetic
cooling at sub-Kelvin temperatures\cite{pourret}. This is not the
case of Bismuth. In spite of its much larger Nernst coefficient, it
does not qualify as a suitable thermomagnetic material. Since the
magnetoresistance is large,  $ZT_{\epsilon}$ remains very small.
This can be seen in Fig. 4, which compares the field-dependence of
$N$ and $ZT_{\epsilon}$ in Bismuth and in PrFe$_{4}$P$_{12}$. As a
consequence of the lightness of carriers in Bismuth,
$\omega_{c}=\frac{eB}{m^{*}}$ is large and the magnetic field
induces a huge decrease in electric conductivity. Meanwhile, a large
heat conductivity is maintained by phonons and the WF law is not
relevant. In contrast with Bismuth, the large Nernst coefficient of
PrFe$_{4}$P$_{12}$ is mostly due to the smallness of $\epsilon_{F}$.
Electrons are heavy, $\omega_{c}$ is not large and the system keeps
its metallic behavior in a magnetic field. This is the fundamental
reason behind its sizeable $ZT_{\epsilon}$ ($\simeq$0.2 at 4 T and 1
K). Such heavy-fermion semi-metals emerge from our analysis as
promising thermomagnetic materials at Kelvin temperatures.
\begin{figure}
\resizebox{!}{0.3\textwidth} {\includegraphics{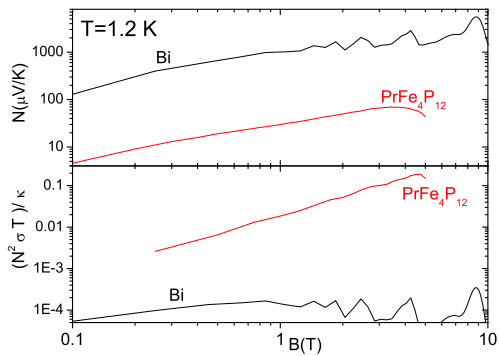}}
\caption{\label{fig4}a) The  Nernst signal (a) and the the
thermomagnetic figure of merit (b) in Bi and in PrFe$_{4}$P$_{12}$
as a function of magnetic field at T= 1.2K.}
\end{figure}

In comparison with its Peltier counterpart, Ettingshausen cooling
presents the obvious drawback of requiring a magnetic field.
However, there are reasons to suspect that it may prove to be
promising. First, contrary to the Seebeck effect and as argued
above, the Nernst effect is expected to scale with the
mean-free-path. Therefore, the purification of the selected
candidate can enhance its thermoelectric performance. Moreover, the
geometry of the Ettingshausen effect allows the design of an
infinite-stage refrigerator\cite{obrien}. To be widely used, such a
cooler should work with permanent magnets which can currently
produce fields of the order of 1 T.

In summary, we studied the Nernst effect in Bismuth and argued that
the electrons present a large Nernst response whenever their Fermi
energy is low, their cyclotron frequency large and their scattering
time long. Moreover, when they are heavy enough, a metallic behavior
is maintained in presence of a magnetic field and thermomagnetic
refrigeration at low temperatures becomes possible. We thank Y.
Nakajima, A. Pourret and M. Nardone for precious assistance, C.
Proust for useful discussions and B. C. Sales for informing us on
the large Nernst effect in Bismuth. This work was supported in
France by the ICENET project (ANR) and in Brazil by CNPq and FAPESP.

\end{document}